\documentclass[aps,prc,preprint,superscriptaddress,showkeys]{revtex4-2}
\usepackage{graphicx}
\usepackage{dcolumn}
\usepackage{bm}
\usepackage{amsmath}
\begin{document}

\title{Theoretical analysis of the reactions induced by interaction of $^{6}$Li with nuclei $^{3}$H and $^{3}$He}
\author{Yu. A. Lashko}
\email{ylashko@gmail.com}
\affiliation{National Institute for Nuclear Physics, Padova Division, 35131 Padova, Italy}
\affiliation{Bogolyubov Institute for Theoretical Physics, 03143 Kyiv, Ukraine}
\author{V. S. Vasilevsky}
\email{vsvasilevsky@gmail.com}
\affiliation{Bogolyubov Institute for Theoretical Physics, 03143 Kyiv, Ukraine}
\author{V. I. Zhaba}
\email{viktorzh@meta.ua}
\affiliation{Bogolyubov Institute for Theoretical Physics, 03143 Kyiv, Ukraine}

\date{\today}

\begin{abstract}
We determine cross sections and astrophysical S-factors of the reactions generated in collisions between $^{6}$Li and $^{3}$H, and $^{6}$Li and $^{3}$He. A microscopic three-cluster model is employed to study the dynamics of reactions occurring in the mirror nuclei $^{9}$Be and $^{9}$B. In a previous study [Phys. Rev. C {\bf 109}, 045803 (2024)], this model was successfully applied to investigate the resonance structure of $^{9}$Be and $^{9}$B, as well as reactions induced by the interaction of deuterons with $^7$Li and $^7$Be. A fairly good agreement between theoretical results and available experimental data was achieved. To the best of our knowledge, appropriate experimental data for the astrophysical S-factors of the reactions generated by the interaction of $^6$Li with $^3$H and with $^3$He are currently unavailable. Thus, our results can serve as a guideline for future experimental efforts.

\end{abstract}

\maketitle

\section{Introduction}

In this paper we will study reactions induced by interaction of $^{6}$Li with $^{3}$H 
\begin{equation}
^{6}\text{Li}+^{3}\text{H}=^{7}\text{Li}+d,\quad^{6}\text{Li}+^{3}
\text{H}=^{8}\text{Be}+n,\quad^{6}\text{Li}+^{3}\text{H}=^{5}\text{He}
+^{4}\text{He}\label{eq:00R1}
\end{equation}
and $^{6}$Li with $^{3}$He
\begin{equation}
^{6}\text{Li}+^{3}\text{He}=^{7}\text{Be}+d,\quad^{6}\text{Li}+^{3}
\text{He}=^{8}\text{Be}+p,\quad^{6}\text{Li}+^{3}\text{He}=^{5}\text{Li}
+^{4}\text{He.}\label{eq:00R2}%
\end{equation}
The first reaction in list (\ref{eq:00R1}) leads to
the synthesis of $^{7}$Li nuclei, while all other reactions in (\ref{eq:00R1})
and (\ref{eq:00R2})  result in the burning of $^{6}$Li nuclei.

Reactions (\ref{eq:00R1}) and (\ref{eq:00R2}) are studied within the framework of a
microscopic multi-channel model advanced in Refs.\cite{PhysRevC.109.045803, 2024FBS....65...14L}. 
In Ref. \cite{PhysRevC.109.045803}, we investigated reactions generated by the collision of deuterons with the nuclei $^{7}$Li and $^{7}$Be. 
These reactions are pertinent to the cosmological lithium problem, which stimulates theoretical and experimental studies of reactions wherein lithium isotopes are synthesized and burned. Given that these processes occurred in the early Universe at relatively low temperature, special attention is paid to these reactions within a low  energy range, specifically, less than 100 keV in the center-of-mass reference frame for the incident nuclei.

Some of the reactions (\ref{eq:00R1}) and (\ref{eq:00R2}) have been the subject of experimental investigations as referenced in studies such as \cite{1956PhRv..104.1064S, 1971NuPhA.173..273Z, 1980PhRvC..22.1406E, 4331412, 2014NDS...120..184P, 2016JPhCS.665a2006P}. Among these, the pioneering study by \cite{1956PhRv..104.1064S} notably expands on the limited data available on low-energy scattering between $^3$He and $^6$Li. This research reported the measurement of the cross-section for the $^6$Li($^3$He,p)$^8$Be reaction across the $^3$He energy spectrum from 1 to 5 MeV. It presented excitation functions for the reactions leading to both the ground state ($0^+$) and the first excited state ($2^+$) of $^8$Be, measured at laboratory angles of 0° and 150°. The analysis revealed two resonances within the compound nucleus $^9$B, occurring approximately 1 MeV and 2 MeV above the $^6$Li + $^3$He threshold, which were particularly pronounced in the production of protons leading to the $2^+$ state of $^8$Be.

In the experiment by Gould and Boyce \cite{GouldBoyce1976}, the cross-section of the $^6$Li($^3$He,p)$2\alpha$ reaction was measured across a range of bombarding energies from 3 to 6 MeV. Their findings indicate that the total cross-sections for the $^6$Li($^3$He,p)$^8$Be$(0^+)$ reaction remain relatively flat from 2 to 10 MeV. Conversely, observations of the $^6$Li($^3$He,p)$^8$Be$(2^+)$ reaction suggest the presence of a broad peak in the cross-section around 4 MeV. By applying an s-wave Gamow extrapolation to cross-sections below 1 MeV, combined with experimental data \cite{1956PhRv..104.1064S} at 1 MeV, the researchers estimated the astrophysical S-factors for these reactions, yielding values of 13.5 MeV·b for the ground state of $^8$Be and 33 MeV·b for its $2^+$ excited state. 

In the study \cite{1980PhRvC..22.1406E}, the authors reported on both total and differential cross-sections for the $^6$Li($^3$He,p)$^8$Be reaction leading to the ground state, the $2^+$ state, and the 16.63- and 16.92-MeV states in $^8$Be. Additionally, they provided preliminary data on the deuterons produced in the $^6$Li($^3$He,d) reaction targeting both the ground and first-excited states of $^8$Be. These measurements were conducted over a range of incident $^3$He energies from 0.5 to 1.85 MeV.

Their findings indicate that, for proton energies exceeding 1 MeV, the energy spectrum is predominantly influenced by $\alpha-\alpha$ interactions within the 0$^+$ ground and 2$^+$ excited states of $^8$Be, with a notable emphasis on the $^6$Li($^3$He,p)$^8$Be$(2^+)$ reaction. The authors also noted that the generation of protons and $\alpha$ particles in the $^6$Li+$^3$He reactions primarily occurs through sequential reactions via the established states of $^8$Be and $^5$Li, with minimal contributions from direct three-body breakups.

Furthermore, the research presents cross-sections for the $^6$Li($^3$He,d)$^7$Be reaction within a constrained energy range of 1.2 to 1.6 MeV for the bombarding $^3$He. These findings are in a good agreement with earlier results from Barr and Gilmore \cite{Barr1965}, who employed an acstivation technique to study this reaction. 
Additionally, reaction-rate parameters for the different branches of the $^6$Li($^3$He,$p$) reaction were determined by extrapolating to zero energy from the data point at the lowest measured energy of 0.5 MeV.

A different extrapolation procedure was used in \cite{Voronchev2003} to calculate astrophysical S-factors of the reaction $^6$Li($^3$He,p)$^8$Be at energies below 1 MeV leading to a production of $^8$Be nuclei. However, only highly-excited states 16.63 and 16.92 MeV of $^8$Be were considered.

In \cite{2014NDS...120..184P}, total cross-section data for the $^6$Li($^3$He,p)$^8$Be reaction, as detailed in \cite{1980PhRvC..22.1406E}, were analyzed. These data cover $^3$He energies ranging from 0.66 MeV to 5.00 MeV, with the analysis averaging the final states across the excited states of the $p+^8$Be quasi-two-body system. Additionally, total cross-section data for the $^6$Li($^3$He,d)$^7$Be reaction, referenced in \cite{4331412}, were examined for $^3$He energies from 0.42 MeV to 4.94 MeV. This comprehensive analysis employed the multichannel, two-body unitary R-matrix method to interpret the reactions across these energy ranges.

The R-matrix analysis presented in \cite{2014NDS...120..184P} identified a resonance structure that notably diverges from the findings published in \cite{Tilley2004} (refer to the table below). This discrepancy might stem from the fact that their analysis incorporates data for the energy range E($^3$He) $<$ 3.0 MeV through a multichannel, unitary R-matrix evaluation, whereas a single-level R-matrix parametrization was used in \cite{Tilley2004}.
\begin{ruledtabular} 
\begin{table}[htbp] \centering
\caption{R-matrix fit of resonance states in $^9$B, with energies measured relative to $^6$Li$+^3$He decay threshold}
\begin{tabular}[c]{cccccc}
\multicolumn{3}{c}{R-matrix analysis \cite{2014NDS...120..184P}} & \multicolumn{3}{c} {Compilation \cite{Tilley2004}}\\\hline
$J^{\pi}$ & $E$, MeV & $\Gamma$, keV & $J^{\pi}$ & $E$, MeV $\pm$ keV & $\Gamma$, MeV $\pm$ keV\\\hline
$1/2^-$ & -0.1369 & 768 &  & -0.5783 $\pm$ 25 & 180 $\pm$ 16\\
$1/2^-$ & 0.5109 & 0.14 & ($5/2^+$) & 0.1077 $\pm$ 100 & \\
$5/2^-$ & 0.5989 & 871.63 & $1/2^-$ & 0.4737 $\pm$ 4 & 22 $\pm$ 5 \\
$3/2^-$ & 0.6785 & 147.78 &  & 0.5877 $\pm$ 25 & 120 $\pm$ 40 \\
$5/2^+$ & 1.0631 & 33.33 & ($7/2^+$) & 0.9377 $\pm$ 100 & \\
$7/2^+$ & 1.2339 & 2036.21 &  & 1.0347 $\pm$ 10  & 71 $\pm$ 8 \\
$3/2^-$ & 1.2454 & 42.52 &  &  & \\
$3/2^+$ & 1.4459 & 767.11 &  &  & \\
$1/2^+$ & 1.8206 & 5446.32 &  &  & \\
$1/2^-$ & 2.0749 & 10278.41 &  &  & \\
$3/2^-$ & 3.0069 & 1478.22 &  &  & \\
\end{tabular}
\label{Rmatrix}
\end{table}
\end{ruledtabular}

In \cite{LeMere1980}, the $^6$Li+$^3$He system was examined using the resonating-group method under the single-channel approximation. In this approach, the $^6$Li nucleus was treated as a cluster comprising two correlated nucleons within the non-closed 1p shell. The Minnesota potential, selected for its nucleon-nucleon interactions, featured an exchange mixture parameter $u=0.95$.

The study analyzed phase shifts for the $^6$Li+$^3$He scattering within the channel-spin states $S=1/2$ and $S=3/2$, considering orbital angular momentum $L\leq7$. The authors identified numerous bound and sharp resonance levels in the energy region below approximately 10 MeV. Notably, a distinct sharp resonance level with quantum numbers $L^\pi=2^+$ and spin $S=1/2$ was located just 0.025 MeV above the $^6$Li+$^3$He threshold. The authors emphasized the significance of incorporating the $d+\alpha$ structure of $^6$Li and considering the $p+^8$Be cluster configuration to enhance the accuracy of their calculations.

The data concerning the cross-section for tritium interaction with $^6$Li is notably scarce. Specifically, within the low-energy range below 2 MeV, reference can be made to the work conducted by Voronchev et al. \cite{Voronchev2000}, where the authors evaluated the cross-sections for the $^6$Li(t, d)$^7$Li($1/2^-$) reaction employing a cluster folding model. This evaluation utilized a similar extrapolation method to the one they applied for calculating the astrophysical S-factors of the reaction $^6$Li($^3$He,p)$^8$Be in their study mentioned above \cite{Voronchev2003}. The extrapolation of the energy dependence for the $^6$Li(t, d)$^7$Li($1/2^-$) cross-sections up to approximately 2 MeV, where experimental data begin to emerge, used nuclear data from Abramovich et al. \cite{Abramovich1991}, supplemented by experimental findings from  \cite{Abramovich1984}. The cross-sections for the $^6$Li(t, d)$^7$Li($1/2^-$) reaction, as discussed in \cite{Voronchev2003}, demonstrate a steady and notable increase with energy reaching up to 1 MeV.

Besides, Voronchev et al. \cite{Voronchev2003} compared their results with prior data derived using traditional Gamov extrapolation methods. They observed that the values from  \cite{Abramovich1991} were considerably lower, by about 2.5 to 3.5 times, in the sub-barrier region. However, it is pertinent to note that the study by Voronchev et al. \cite{Voronchev2003} omitted the potential impact of excited states in $^9$Be near the $^6$Li+t threshold, specifically around 17.7 MeV, which could influence the $^6$Li+t reaction cross-section at lower energies.

In terms of the $^6$Li(t, d)$^7$Li($3/2^-$) cross-section, Abramovich et al. \cite{Abramovich1991} provided assessments for an energy range from 0.3 MeV to 4.12 MeV, drawing on data from the inverse reaction $^7$Li(d, t)$^6$Li reported by Macklin \cite{Macklin1955}. Their findings indicated that the cross-section leading to the ground state of $^7$Li is roughly twice that of the reaction leading to its excited state, with a peak observed near 3 MeV.

Total cross-sections for neutron production in the $^6$Li(t, n)$^8$Be reaction for triton energies spanning from 0.358 MeV to 2.123 MeV were measured by Valter \cite{Valter1962}. A notable resonance in neutron yield was identified at a triton energy of 1.875 MeV, corresponding to an excitation level of 18.936 MeV in $^9$Be. Abramovich et al. \cite{Abramovich1991} estimated the S-factor for the $^6$Li(t, n)$^8$Be reaction to be $S(0) = (1.78 +0.27)\times10^{-4}$ MeV mb.

In investigating the origins of the $^7$Li problem, Broggini et al. \cite{Broggini2012} examined the nuclear reaction cross sections relevant to Big Bang Nucleosynthesis. They proposed a framework to assess how changes in nuclear reaction rates could influence the primordial abundance of $^7$Li, employing basic nuclear physics concepts. Noting that $^7$Li predominantly forms from $^7$Be through electron capture, they analyzed the effects of different $^7$Be destruction channels. Their findings indicated that only the reactions involving $^7$Be+$d$ and $^7$Be+$^4$He have a substantial potential to reduce $^7$Li abundance, with maximum reductions of 40\% and 25\% respectively. They hypothesized the existence of an unconfirmed resonance in the $^9$B nucleus, about 150 keV above the $^7$Be+$d$ threshold, which could significantly influence $^7$Li levels but would not completely solve the lithium discrepancy. Additionally, they pointed out that the reaction $^7$Be+$d\rightarrow^6$Li$+^3$He might enhance $^6$Li production in the early universe.

Therefore, it is evident that additional theoretical and experimental exploration into the reactions facilitated by $^6$Li interactions with tritium and $^3$He remains imperative. Such investigations are crucial to understand the resonance behavior near the $^6$Li+t and $^6$Li+$^3$He thresholds in $^9$Be and $^9$B nuclei, as well as their impact on the formation of $^7$Li and the burning of $^6$Li nuclei.

\section{Multi-channel model}
In this section we briefly present main ideas of  the  model which was explained in more details in Refs. \cite{PhysRevC.109.045803, 2024FBS....65...14L}.

In the present paper, we use a microscopic multi-channel cluster model which
 takes into account a major part of binary channels in $^{9}$Be and
$^{9}$B. For the $^{9}$Be nucleus they are%
\begin{equation}
^{6}\text{Li}+^{3}\text{H},\quad^{7}\text{Li}+d,\quad^{8}\text{Be}+n,\quad
^{5}\text{He}+^{4}\text{He}\label{eq:001}%
\end{equation}
and in $^{9}$B nucleus they involve%
\begin{equation}
^{6}\text{Li}+^{3}\text{He},^{7}\text{Be}+d,^{8}\text{Be}+p,^{5}\text{Li}%
+^{4}\text{He.}\label{eq:002}%
\end{equation}
These channels are important for describing reactions of burning and synthesis of $^{6,7}$Li.

Our focus is on the energy region near the $^6$Li+$^3$H threshold in $^9$Be and $^6$Li+$^3$He in $^9$B, situated between two  different three-cluster thresholds of $^9$Be and $^9$B that generate the aforementioned binary channels.

Consequently, it is natural to describe $^{9}$Be as a system of two coupled three-cluster configurations:
\begin{equation}
^{4}\text{He}+^{4}\text{He+n, }^{4}\text{He}+^{3}\text{H}+d,\label{eq:003}.
\end{equation}
Similarly, for $^{9}$B, we employ the mirror three-cluster configurations:
\begin{equation}
^{4}\text{He}+^{4}\text{He+p, }^{4}\text{He}+^{3}\text{He}+d.\label{eq:004}%
\end{equation}

Within this approximation, the wave function $\Psi_{E,J}$, which describes the structure of the $^{9}$B and $^{9}$Be compound nuclei in the state with total angular momentum $J$ and energy $E$, as well as various binary reactions proceeding through these compound nuclei, becomes a multi-component function:
\begin{equation}
\Psi_{E,J}=\left(
\begin{array}
[c]{c}%
\Psi_{1}^{\left(  E,J\right)  }\\
\Psi_{2}^{\left(  E,J\right)  }\\
\Psi_{3}^{\left(  E,J\right)  }\\
\Psi_{4}^{\left(  E,J\right)  }.%
\end{array}
\right)  \label{eq:020A}%
\end{equation}

Each component $\Psi_{k}^{\left(E,J\right)}$ of the three-cluster system is represented as an antisymmetrized product combining the intrinsic wave functions of an $s$-shell cluster with mass number $A_1\leq4$ and a two-cluster subsystem with mass number $A_2=A_{21}+A_{22}$ ($A_{21}, A_{22}\leq4$), along with the wave function for their relative motion:
\begin{equation}
\Psi_{k}^{\left(  E,J\right)  }=\widehat{\mathcal{A}}\left\{  \Phi_{1}\left(
A_{1},S_{1}\right)  \psi_{2}\left(  \mathcal{E}_{2},A_{2},j_{2}\right)
\varphi_{E-\mathcal{E}_{2},c_{k}}\left(  \mathbf{y}_{k}\right)  \right\}  _{J}\label{eq:020B}%
\end{equation}
The wave function (\ref{eq:020B}) models the two-cluster dynamics within the considered three-cluster systems of $^9$Be or $^9$B for a specific binary channel denoted by the index $k$. Here, $k = 1, 2, 3, \text{ or } 4$, corresponds to one of the four channels listed in either Eq.~(\ref{eq:001}) for $^{9}$Be or Eq.~(\ref{eq:002}) for $^{9}$B. 

Index $c_{k}$ in Eq. (\ref{eq:020B}) is a multiple index that unambiguously enumerates all channels of a given compound nucleus compatible with the selected cluster partition $k$. A Jacobi vector $\mathbf{y}_{k}$ determines the relative position of an $s$-shell cluster consisting of $A_1$ nucleons and a two-cluster subsystem consisting of $A_2$ nucleons, which is in the state with energy $\mathcal{E}_{2}$. The wave function $\varphi_{E-\mathcal{E}_{2},c_{k}}\left(  \mathbf{y}_{k}\right)  $ describes their relative motion.

Note that each component $\Psi_{k}^{(E,J)}$ describes either the decay of a compound system (if this channel is an exit channel) or the formation of a compound system (if this channel is an entrance channel).

The first function in Eq. (\ref{eq:020B}),
$\Phi_{1}\left(  A_{1},S_{1}\right)  $, denotes the shell-model wave function
of a $s$-shell cluster ($p$, $n$, $d$, $^{3}$H, $^{3}$He, $^{4}$He), reflecting its intrinsic structure. The second function, $\psi_{2}\left(  \mathcal{E}_{2},A_{2},j_{2}\right),$
characterizes the internal dynamics of two-cluster subsystems such as $^{5}$He, $^{5}$Li, $^{6}$Li, $^{7}$Li, $^{7}$Be, and $^{8}$Be, emerging as solutions to the respective two-cluster Schr\"{o}dinger equation. 

Prior to deducing equations for the wave function $\varphi_{E-\mathcal{E}_{2},c_{k}}(\mathbf{y}_{k})$ and solving them, it is essential to define the wave function of a two-cluster subsystem, $\psi_{2}(\mathcal{E}_{2},A_{2},j_{2})$, which describes the interaction of two clusters in both bound and pseudo-bound states. The structure of the wave function $\psi_{2}(\mathcal{E}_{2},A_{2},j_{2})$ mirrors that of the wave function defined in Eq. (\ref{eq:020B}).
Therefore, for a cluster partition defined by $A_{2}=A_{21}+A_{22}$, it can be expressed as an antisymmetrized product of the intrinsic cluster wave functions $\Phi_{21}$ and $\Phi_{22},$ and a wave function $\chi$ describing their relative motion
\begin{equation}
\psi_{2}(\mathcal{E}_{2},A_{2},j_{2}) = \widehat{\mathcal{A}}\left\{ \left[ \Phi_{21}(A_{21},S_{21}) \Phi_{22}(A_{22},S_{22}) \right]_{S_{2}}\chi_{\mathcal{E}_{2},\lambda}(\mathbf{x}_{k}) \right\}_{j_{2}},
\end{equation}
where $\mathbf{x}_{k}$ represents the vector specifying the distance between clusters $A_{21}$ and $A_{22}$, while $\lambda$ and $j_{2}$ denote the orbital and angular momenta characterizing the relative motion of these clusters.
In our model, both clusters $A_{21}$ and $A_{22}$ belong to $s$-shell clusters possessing zero intrinsic orbital angular momentum. Consequently, their intrinsic angular momenta are solely determined by their spins, denoted as $S_{21}$ and $S_{22}$, respectively.

Wave function ${\chi}_{\mathcal{E}_{2},\lambda}\left(  \mathbf{x}_{k}\right)$
of the relative motion between clusters $A_{21}$ and $A_{22}$ in a two-cluster subsystem $A_2$
is obtained by solving the two-cluster Schrödinger equation. The solutions to this equation yield the energy $\mathcal{E}_{2}$ associated with bound and pseudo-bound states of the two-cluster subsystem. This energy also sets the threshold energy for a binary channel $A_1+A_2$ within the compound three-cluster system $A=A_1+(A_{21}+A_{22})$.

To obtain wave function $\varphi_{E-\mathcal{E}_{2},c_{k}}(\mathbf{y}_{k})$, which describes the relative motion of the two-cluster subsystem $A_2=A_{21}+A_{22}$ and a remaining cluster $A_1$,  
and other crucial parameters of the three-cluster compound system, a set of coupled equations can be derived as follows:

\begin{equation}
\sum_{\widetilde{c}}\left[  \left\langle c\left\vert \widehat{H}\right\vert
\widetilde{c}\right\rangle -E\left\langle c|\widetilde{c}\right\rangle
\right]  \varphi_{E-\mathcal{E}_{2},c_{k}}=0\label{eq:021}%
\end{equation}
by employing a projection operator that maps the three-cluster system onto the binary channel $A=A_1+A_2$:
\begin{equation}
\widehat{\mathcal{P}}_{c}=\widehat{\mathcal{A}}\left\{  \Phi_{1}\left(
A_{1},S_{1}\right)  \psi_{2}\left(  \mathcal{E}_{2},A_{2},j_{2}\right)
\delta\left(  \mathbf{y}_{c}^{\prime}-\mathbf{y}_{c}\right)  \right\}
.\label{eq:022}%
\end{equation}
With this operator, we obtain
\begin{eqnarray}
\left\langle c\left\vert \widehat{H}\right\vert \widetilde{c}\right\rangle & = \left\langle \widehat{\mathcal{P}}_c \left\vert \widehat{H}\right\vert \widehat{\mathcal{P}}_{\widetilde{c}}\right\rangle, \label{eq:023} \\
\left\langle c|\widetilde{c}\right\rangle & = \left\langle \widehat{\mathcal{P}}_c |\widehat{\mathcal{P}}_{\widetilde{c}}\right\rangle, \label{eq:023A}
\end{eqnarray}
where $\widehat{H}$ is the Hamiltonian of the three-cluster system $A_1 + A_{21} + A_{22} = A_1 + A_2$, and integration is performed over all spatial, spin, and isospin variables.

It was shown by J. Wheeler \cite{1937PhRv...52.1083W, 
 1937PhRv...52.1107W},  that the matrix elements of the Hamiltonian $\left\langle c\left\vert \widehat{H}\right\vert \widetilde{c}\right\rangle$
are integro-differential operators, and the matrix elements of the unit operator $\left\langle c|\widetilde{c}\right\rangle$ are integral operators. A similar operator to the operator (\ref{eq:022}) can be constructed for a two-cluster subsystem $A_2=A_{21}+A_{22}$, and  similar integro-differential equation can be derived for the wave function $\chi_{\mathcal{E}_{2},\lambda}\left(\mathbf{x}_{k}\right),$ describing the intrinsic structure of the two-cluster subsystem.

Now we can specify the composite index $c_{k}$. For a given value of the total angular momentum $J$ and parity $\pi$, the index $c_{k}$ encompasses the following quantum numbers:
\begin{equation}
c_{k} = \{k, \lambda, j_{1}, l, j_{2}\}. \label{eq:qn1}
\end{equation}
Here, $l$ and $j_{2}$ are the orbital and total angular momenta of the "target" (two-cluster subsystem), respectively, while $\lambda$ and $j_{1}$ represent the orbital and total angular momenta of the relative motion between the "projectile" and the "target".

For the numerical realization of our model, we employ square-integrable bases: a basis of Gaussian functions for the two-cluster subsystems and a basis of oscillator functions for the compound system. These bases are used to expand the wave functions $\chi_{\mathcal{E}_{2},\lambda}(\mathbf{x}_{k})$ and $\varphi_{E-\mathcal{E}_{2},c_{k}}(\mathbf{y}_{k})$, respectively. As a result, we must numerically solve a system of linear algebraic equations (for details, see Ref. \cite{PhysRevC.109.045803} and the references therein).

\section{Results}

In this section, we briefly outline the selection of the main input parameters and provide some details of our model. The primary input parameter for any microscopic model is the nucleon-nucleon (NN) potential. We have chosen the Minnesota potential (MP) \cite{1977NuPhA.286...53T,  1970NuPhA.158..529R} for all calculations. Following the selection of the NN potential, the only model parameter we need to determine is the oscillator radius, also known as the oscillator length, $b$. 

Traditionally, this oscillator length is selected to minimize the energy of the lowest three-cluster threshold—specifically $^{4}$He$+^{4}$He$+n$ in $^{9}$Be and $^{4}$He$+^{4}$He$+p$ in $^{9}$B. This approach effectively optimizes the internal structure of the alpha particle, which is a critical component in both three-cluster configurations (\ref{eq:003}) in $^{9}$Be and (\ref{eq:004}) in $^{9}$B. The minimum energy of the alpha particle is achieved at $b=1.2846$ fm.

The MP is characterized by an exchange parameter $u$, which is adjustable to better align theoretical predictions with experimental data. Following the methodologies outlined in previous studies \cite{PhysRevC.109.045803, 2024FBS....65...14L}, we select $u = 0.95$. This choice aims to closely replicate the experimental ground state energies of $^{7}$Li and $^{7}$Be relative to their respective thresholds ($^{4}$He + $^{3}$H and $^{4}$He + $^{3}$He).
Our calculations yield the ground state energy of $^{6}$Li as $E = -1.29$ MeV, which is slightly below the experimental value of $E = -1.47$ MeV. 

To determine the energies of bound and pseudo-bound states in two-cluster subsystems, we utilize four Gaussian functions. This number of functions guarantees calculations of energies with high precision. Additionally, we employ 150 oscillator functions to describe the interaction between clusters and the structure of compound systems. This quantity of oscillator functions enables us to achieve a unitarity in the scattering S-matrix better than 1\%.
 
The detailed properties of the two-cluster subsystems of $^9$Be and $^9$B at chosen parameters are given in Table 3 of our paper \cite{PhysRevC.109.045803}.

\subsection{Resonance states}

In a previous study \cite{PhysRevC.109.045803}, we calculated the spectrum of resonance states for $^{9}$Be and $^{9}$B around the thresholds of $^{7}$Li+$d$ and $^{7}$Be+$d$, respectively. In Table \ref{Tab:SpectrRS9Bevs9B}, we present a selection of these resonance states that are relevant to the subject of the current paper. Contrary to the approach in Ref. \cite{PhysRevC.109.045803}, the energies of the resonance states in Table \ref{Tab:SpectrRS9Bevs9B} are measured from the thresholds of $^{6}$Li+$^{3}$H and $^{6}$Li+$^{3}$He, respectively. 
Comparing our results with those obtained from R-matrix analysis, we can associate the lowest $1/2^-$ and $3/2^-$ resonances, listed in Table \ref{Tab:SpectrRS9Bevs9B} at energies 0.408 MeV and 0.628 MeV above the $^{6}$Li+$^{3}$He thresholds, with similar resonances found in Table \ref{Rmatrix} at energies 0.5109 MeV and 0.6785 MeV, respectively. These resonances have similar energies, the same angular momenta and parity, but are narrower than ours.
\begin{ruledtabular} 
\begin{table}[htbp] \centering
\caption{Calculated spectrum of resonance states in $^9$B and  $^9$Be, with energies measured relative to the $^6$Li$+^3$He and $^6$Li$+t$ decay threshold}
\begin{tabular}
[c]{cccccc}
\multicolumn{3}{c}{$^{9}$Be} & \multicolumn{3}{c}{$^{9}$B}\\
\hline
$J^{\pi}$ & $E$, MeV & $\Gamma$, MeV & $J^{\pi}$ & $E$, MeV & $\Gamma$,
MeV\\\hline
$\frac{7}{2}^{-}$ & -2.489 & 0.222 & $\frac{7}{2}^{-}$ & -0.685 &
0.116\\
$\frac{1}{2}^{-}$ & -0.787 & 0.680 & $\frac{1}{2}^{-}$ & 0.408 & 1.006\\
$\frac{3}{2}^{-}$ & -0.320 & 1.418 & $\frac{3}{2}^{-}$ & 0.628 & 0.812\\
$\frac{1}{2}^{-}$ & 0.104 & 1.101 & $\frac{1}{2}^{+}$ & 0.668 & 1.867\\
$\frac{3}{2}^{-}$ & 0.239 & 1.610 & $\frac{1}{2}^{-}$ & 0.911 & 1.025\\
$\frac{5}{2}^{-}$ & 0.243 & 1.262 & $\frac{3}{2}^{-}$ & 1.015 & 1.836\\
$\frac{1}{2}^{-}$ & 0.263 & 0.974 & $\frac{5}{2}^{-}$ & 1.033 & 1.585\\
$\frac{1}{2}^{+}$ & 0.278 & 1.238 & $\frac{5}{2}^{-}$ & 1.101 & 1.539\\
$\frac{3}{2}^{-}$ & 0.312 & 1.987 & $\frac{1}{2}^{-}$ & 1.124 & 1.834\\
$\frac{7}{2}^{+}$ & 1.291 & 1.803 & $\frac{7}{2}^{+}$ & 2.509 & 2.377\\
$\frac{5}{2}^{-}$ & 1.374 & 1.735 & $\frac{5}{2}^{-}$ & 2.529 & 1.832\\
$\frac{3}{2}^{-}$ & 1.479 & 2.453 & $\frac{7}{2}^{-}$ & 2.604 & 2.516\\
$\frac{7}{2}^{-}$ & 1.629 & 2.203 & $\frac{3}{2}^{-}$ & 2.757 & 3.482\\
$\frac{7}{2}^{+}$ & 1.710 & 1.827 & $\frac{7}{2}^{-}$ & 2.824 & 2.608\\
\end{tabular}
\label{Tab:SpectrRS9Bevs9B}%
\end{table}%
\end{ruledtabular} 

\begin{figure}[hptb]
\begin{center}
\includegraphics[width=\textwidth]{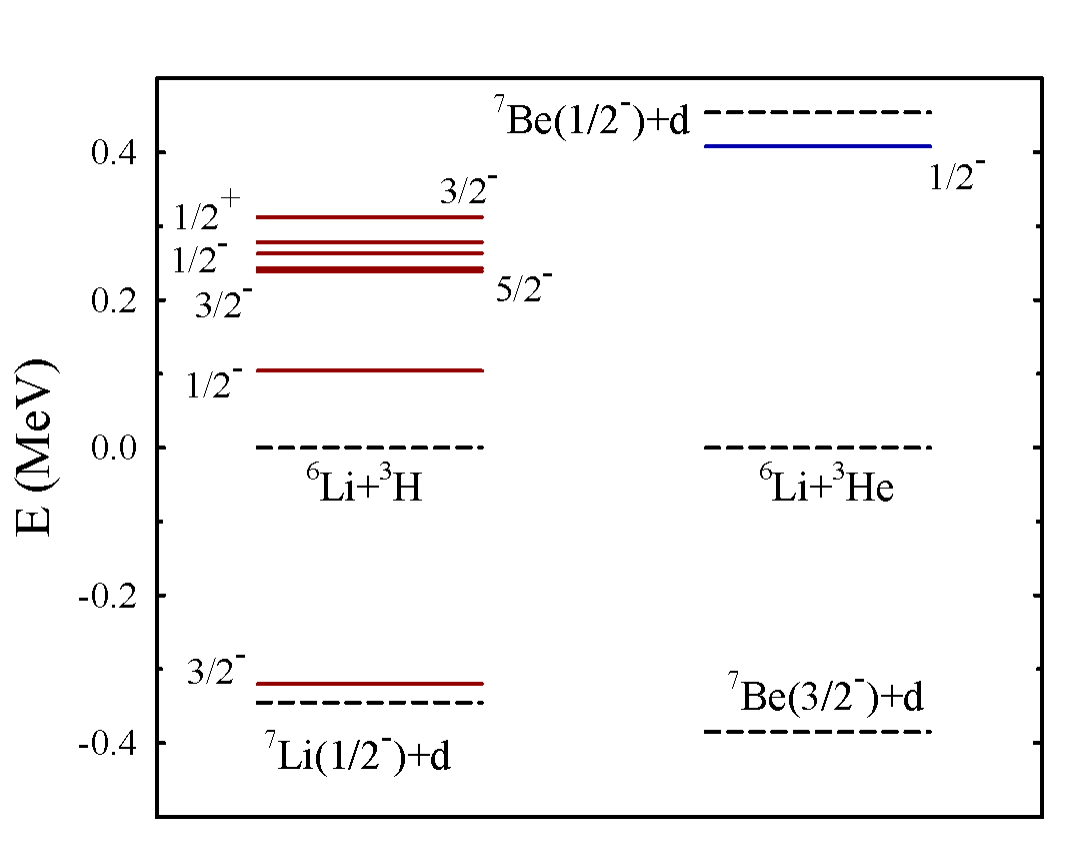}
\caption{Position of resonance states in $^{9}$Be (left) and $^{9}$B (right) around the decay
thresholds $^{6}$Li+$^{3}$H and $^{6}$Li+$^{3}$He, respectively}%
\label{FigResonans9Be9B63}
\end{center}
\end{figure}

In Fig. \ref{FigResonans9Be9B63}, we illustrate the relative positions of the main thresholds in $^{9}$Be and $^{9}$B, which lie close to the thresholds of $^{6}$Li+$^{3}$H and $^{6}$Li+$^{3}$He, respectively. Fig. \ref{FigResonans9Be9B63} also displays several resonance states identified near these thresholds. We anticipate that these resonances may influence the behavior of the astrophysical S-factors arising from the collisions involving $^{6}$Li+$^{3}$H and $^{6}$Li+$^{3}$He.

\subsection{Astrophysical S-factors}
In this section, we analyze astrophysical reactions initiated by the interaction of $^{6}$Li with tritons and $^{3}$He.

\subsubsection{Total and partial S-factors in reactions induced by $^{6}$Li with tritons}

\begin{figure}[hptb]
\begin{center}
\includegraphics[width=\textwidth]{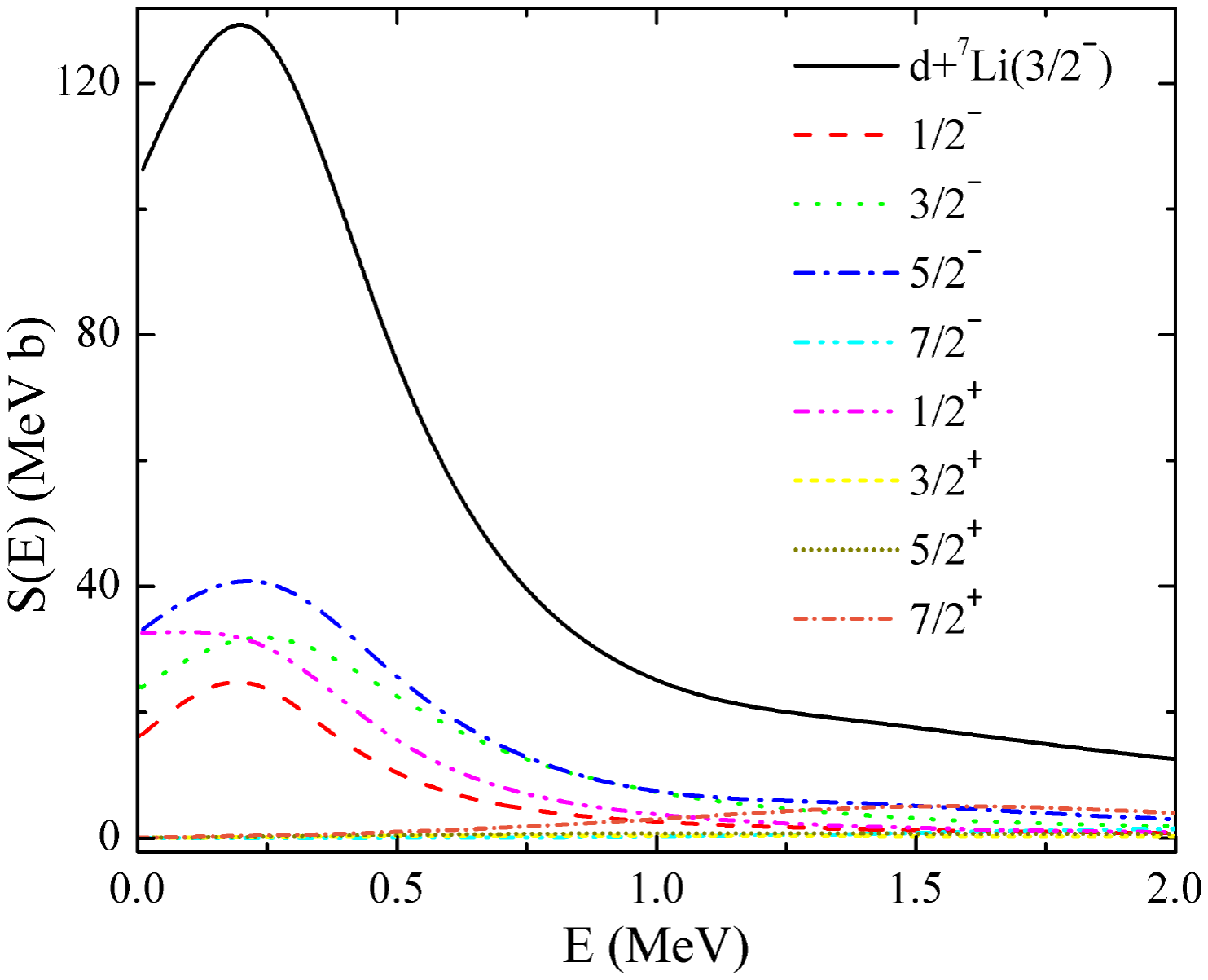}
\caption{Total and partial astrophysical S-factors of the reaction $^{6}
$Li+$^{3}$H=$^{7}$Li(3/2$^{-}$)+$d$ for different $J^{\pi}$ states.}%
\label{Fig:Sfact7Li32MD}
\end{center}
\end{figure}

\begin{figure}[ptb]
\begin{center}
\includegraphics[width=\textwidth]{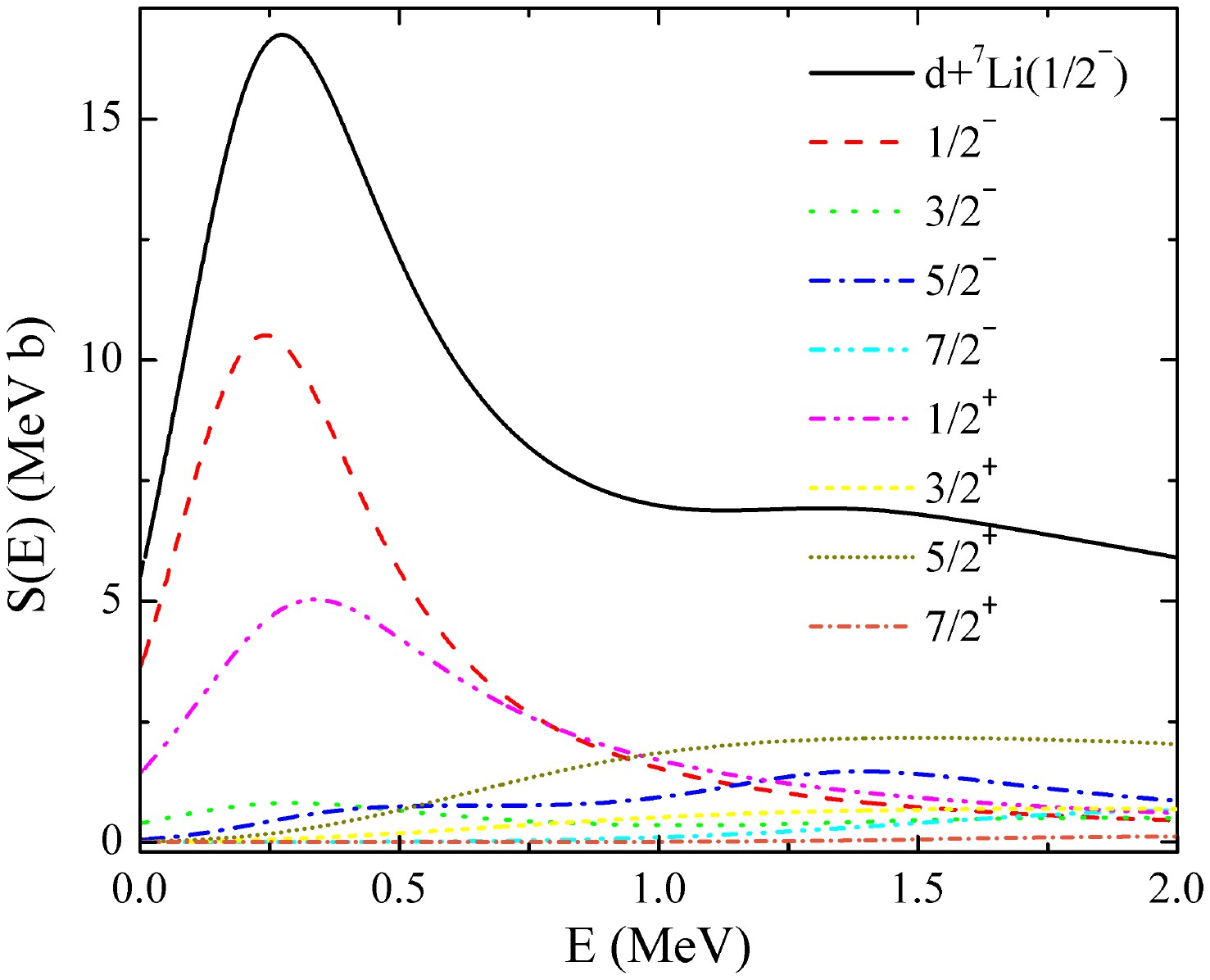}
\caption{Total and partial astrophysical S-factors of the reaction $^{6}
$Li+$^{3}$H=$^{7}$Li(1/2$^{-}$)+$d$ for different $J^{\pi}$ states.}
\label{Fig:Sfact7Li12MD}
\end{center}
\end{figure}

In Figs. \ref{Fig:Sfact7Li32MD} and \ref{Fig:Sfact7Li12MD}, we present the astrophysical $S$-factors for the synthesis reactions of $^{7}$Li, specifically $^{6}$Li+$^{3}$H $\rightarrow$ $^{7}$Li(3/2$^{-}$)+$d$ and $^{6}$Li+$^{3}$H $\rightarrow$ $^{7}$Li(1/2$^{-}$)+$d$. The first reaction predominantly proceeds through the negative parity states 5/2$^{-}$ and 3/2$^{-}$, which are both very close, at approximately 0.24 MeV above the $^{6}$Li+$^{3}$H threshold. Additionally, the 1/2$^{+}$ state, with an energy of $E = 0.278$ MeV, significantly contributes to the peak of the $S$-factor, especially at low-energy region 0$\leq E<$0.3 MeV. It is noteworthy that this state corresponds to a head-on collision between $^6$Li and a triton. 

The reaction synthesizing $^{7}$Li in its ground state, 3/2$^{-}$, occurs nearly eight times more frequently than the reaction producing $^{7}$Li in its excited 1/2$^{-}$ state.  This significant prevalence aligns with the predictions made in Ref. \cite{Abramovich1991}, although they estimated a lower ratio, predicting the reaction $^{6}$Li+$^{3}$H=$^{7}$Li(3/2$^{-}$)+$d$ to be only twice as prevalent. The  reaction $^{6}$Li+$^{3}$H=$^{7}$Li(1/2$^{-}$)+$d$ mainly proceeds via the 1/2$^{-}$ and 1/2$^{+}$ states.
The most prominent peak is formed by the 1/2$^{-}$ resonance state, detected at an energy of 0.263 MeV above the $^{6}$Li+$^{3}$H threshold.  
Conversely, the contribution of this 1/2$^{-}$ resonance state to the reaction $^{6}$Li+$^{3}$H=$^{7}$Li(3/2$^{-}$)+$d$ is relatively small. 

It is interesting to note that the 1/2$^{+}$ resonance state forms the second most important peak in the reaction $^{6}$Li+$^{3}$H=$^{7}$Li(1/2$^{-}$)+$d$, but plays a less significant role in the reaction $^{6}$Li+$^{3}$H=$^{7}$Li(3/2$^{-}$)+$d$.
Overall, the results depicted in Figs. \ref{Fig:Sfact7Li32MD} and \ref{Fig:Sfact7Li12MD} demonstrate a significantly high probability for the synthesis of $^{7}$Li in collisions between $^{6}$Li and $^{3}$H.

\begin{figure}[hptb]
\begin{center}
\includegraphics[width=\textwidth]{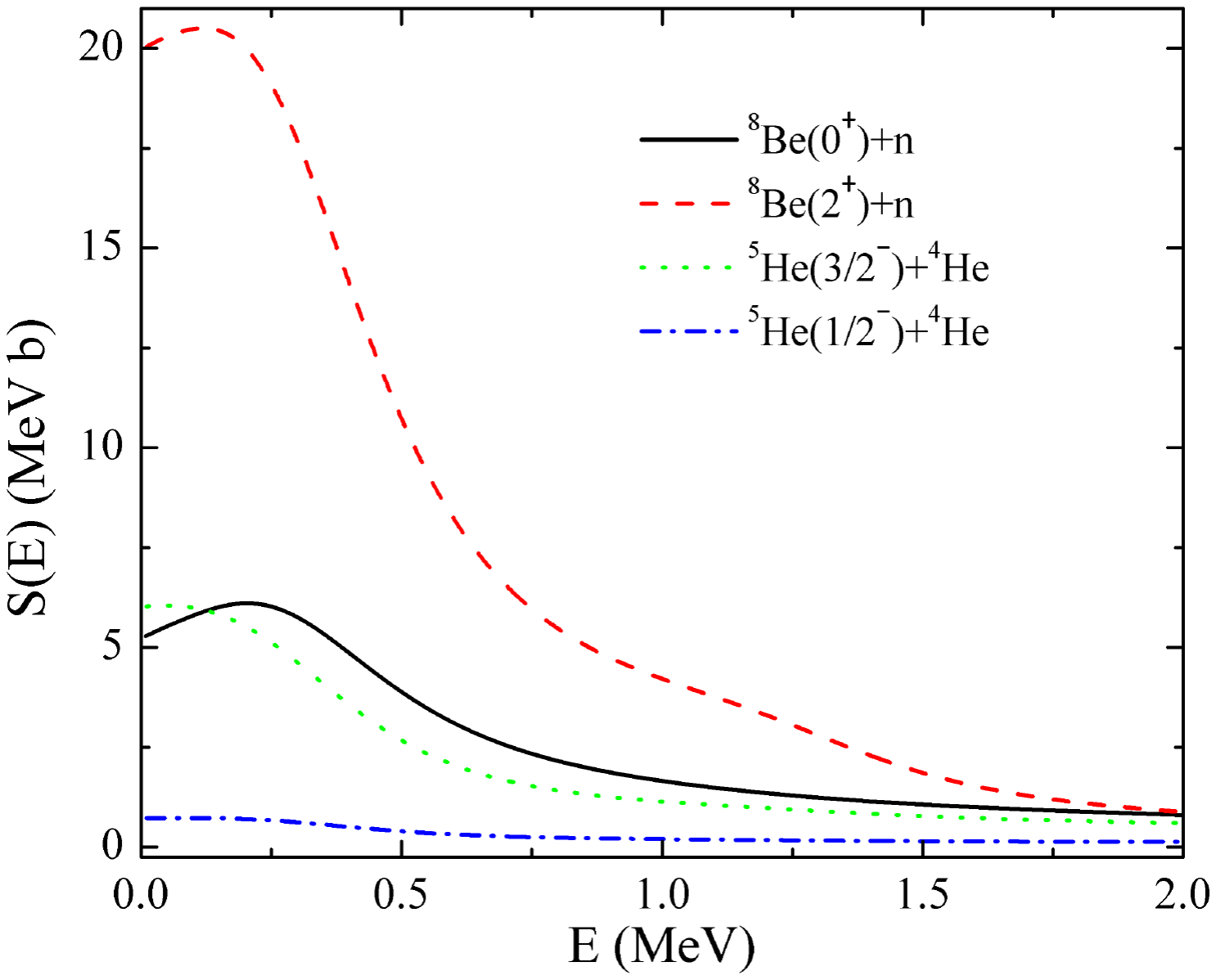}
\caption{Total astrophysical S-factors of the reactions induced by $^{6}$Li+$^{3}$H. Exit channels are listed near the corresponding curves.} 
\label{Fig:Sfacts9Be4}
\end{center}
\end{figure}

The total astrophysical S-factors for reactions with the exit channels $^{8}$Be+$n$ and $^{5}$He+$^{4}$He are illustrated in Fig. \ref{Fig:Sfacts9Be4}. This figure demonstrates that the S-factor for the reaction $^{6}$Li+$^{3}$H $\rightarrow$ $^{8}$Be(2$^{+}$)+$n$ is significantly higher than those for other reactions. However, it remains substantially lower than the S-factor for the synthesis of $^{7}$Li in its ground state via $^{6}$Li+$^{3}$H $\rightarrow$ $^{7}$Li(3/2$^{-}$)+$d$.

\subsubsection{Total and partial S-factors in reactions induced by $^{6}$Li with $^3$He}

\begin{figure}[ptb]
\begin{center}
\includegraphics[width=\textwidth]{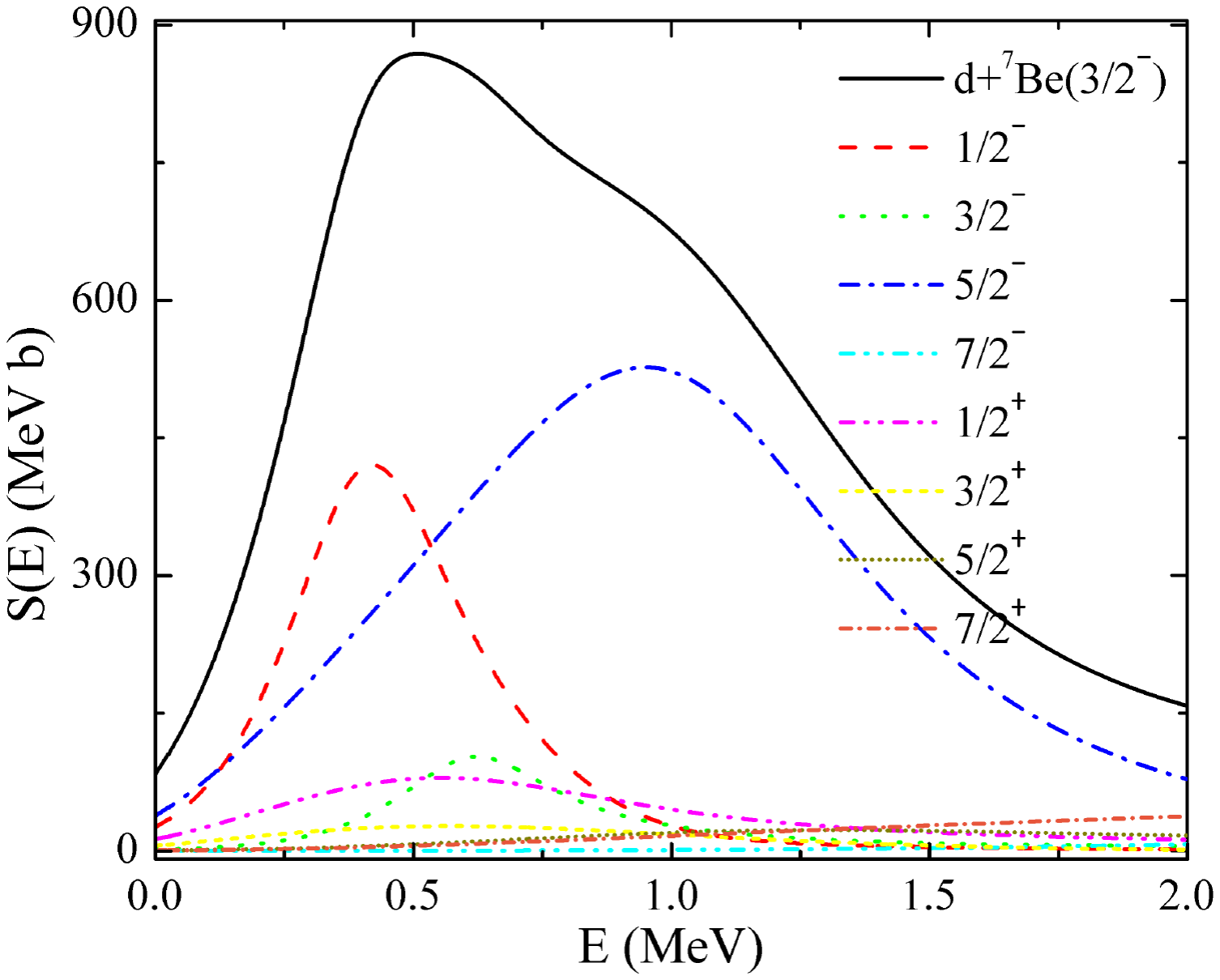}
\caption{Total and partial astrophysical S-factors of the reaction $^{6}$Li+$^{3}$He=$^{7}$Be($3/2^{-}$)+$d$ for different $J^{\pi}$ states.}
\label{Fig:Sfact9B7BeD}
\end{center}
\end{figure}

Fig. \ref{Fig:Sfact9B7BeD} displays the astrophysical $S$-factor for the reaction $^{6}$Li+$^{3}$He $\rightarrow$ $^{7}$Be($3/2^{-})$ + $d$ across different $J^{\pi}$ states. It is evident that the largest $S$-factors are produced by the states 5/2$^{-}$, 1/2$^{-}$, and 3/2$^{-}$. The first peak is generated by the 1/2$^{-}$ resonance state at an energy of $E$=0.408 MeV, while the second peak corresponds to the 5/2$^{-}$ resonance state at $E$=1.033 MeV. Unlike the reactions $^{6}$Li+$^{3}$H $\rightarrow$ $^{7}$Li(3/2$^{-}$)+$d$ and $^{6}$Li+$^{3}$H $\rightarrow$ $^{7}$Li(1/2$^{-}$)+$d$, the 1/2$^{+}$ resonance state, which represents a head-on collision between $^{6}$Li and $^{3}$He, does not contribute significantly to the $S$-factor for the reaction $^{6}$Li+$^{3}$He $\rightarrow$ $^{7}$Be(3/2$^{-}$)+$d$.

\begin{figure}[hptb]
\begin{center}
\includegraphics[width=\textwidth]{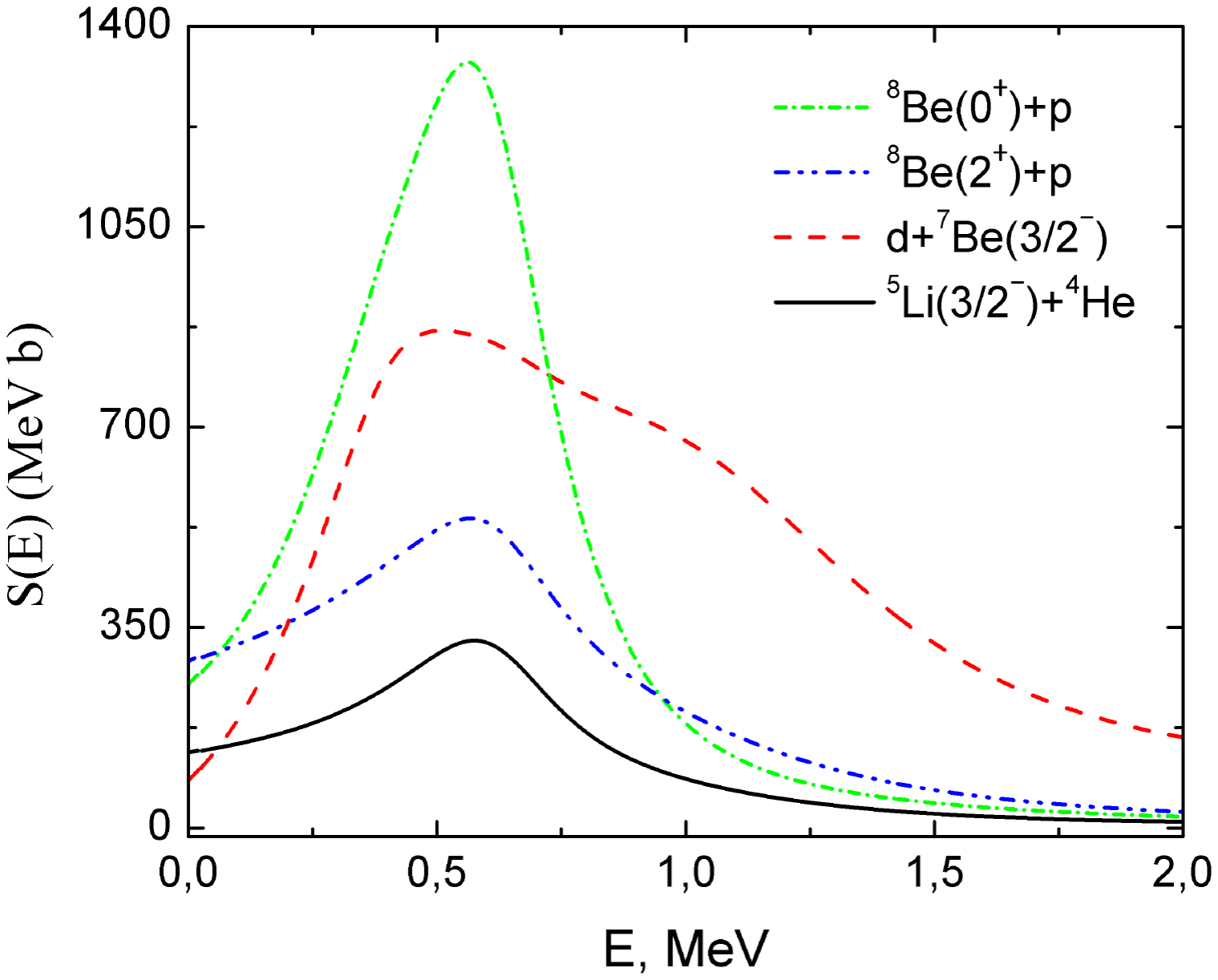}
\caption{Total astrophysical S-factors of the reactions induced by the interaction of $^3$He and $^6$Li. Exit channels are indicated near the corresponding curves.}
\label{Fig:TotalSfact9B}%
\end{center}
\end{figure}

Total astrophysical S-factors for reactions induced by $^{6}$Li+$^{3}$He collisions are displayed in Fig. \ref{Fig:TotalSfact9B}.  
In the low-energy range of 0$\leq E<$0.7 MeV, the S-factor for the reaction $^{6}$Li+$^{3}$He=$^{8}$Be(0$^{+}$)+$p$ dominates, exceeding that of the secondary channel, $^{6}$Li+$^{3}$He=$^{8}$Be(2$^{+}$)+$p$, by 1.5 times at its maximum. At higher energies, the S-factor for the reaction $^{6}$Li+$^{3}$He=$^{8}$Be(2$^{+}$)+$p$ dominates, consistent with the results reported in Refs. \cite{1956PhRv..104.1064S} and \cite{1980PhRvC..22.1406E}. 
Significant peaks in the S-factor for the reaction $^{6}$Li+$^{3}$He=$^{8}$Be(0$^{+}$)+$p$ and the maxima for the S-factors of the reactions $^{6}$Li+$^{3}$He=$^{8}$Be(2$^{+}$)+$p$ and $^{6}$Li+$^{3}$He=$^{5}$Li(3/2$^{-}$)+$^{4}$He are created by the 1/2$^{-}$ resonance state at an energy $E=0.408$ MeV (see Table \ref{Tab:SpectrRS9Bevs9B} and Fig. \ref{FigResonans9Be9B63}).

It is noteworthy that near zero energy, the production of $^8$Be in its $2^+$ excited state becomes slightly more probable than its production in the ground state, somewhat aligning with the predictions made by Gould and Boyce \cite{GouldBoyce1976}. However, the absolute values of S-factors in our calculations are approximately an order of magnitude higher than those estimated by \cite{GouldBoyce1976}. This discrepancy highlights the significant uncertainties noted by Gould and Boyce in low-energy cross-sections and astrophysical S-factors, which they attributed to the limited experimental data available below 1 MeV.

It was anticipated that the 1/2$^{+}$ states would predominantly influence the astrophysical S-factors of reactions driven by collisions between $^{6}$Li and $^{3}$H, as well as $^{6}$Li and $^{3}$He, due to the possibility of head-on collisions in these 1/2$^{+}$ states. Observations confirm this hypothesis for the formation of $^{9}$Be via $^{6}$Li+$^{3}$H collisions, where the 1/2$^{+}$ states significantly contribute to the S-factor. However, in the case of $^{9}$B, formed from the interaction of $^{6}$Li with $^{3}$He, it is the 1/2$^{-}$ state that predominantly enhances the astrophysical S-factors, deviating from the initial expectations.

\subsubsection{S-factors of the reactions $^{7}$Li$+d$ and $^{6}$Li$+t$ with the same exit channel $^{8}$Be$+n$}

\begin{figure}[hptb]
\begin{center}
\includegraphics[width=\textwidth]{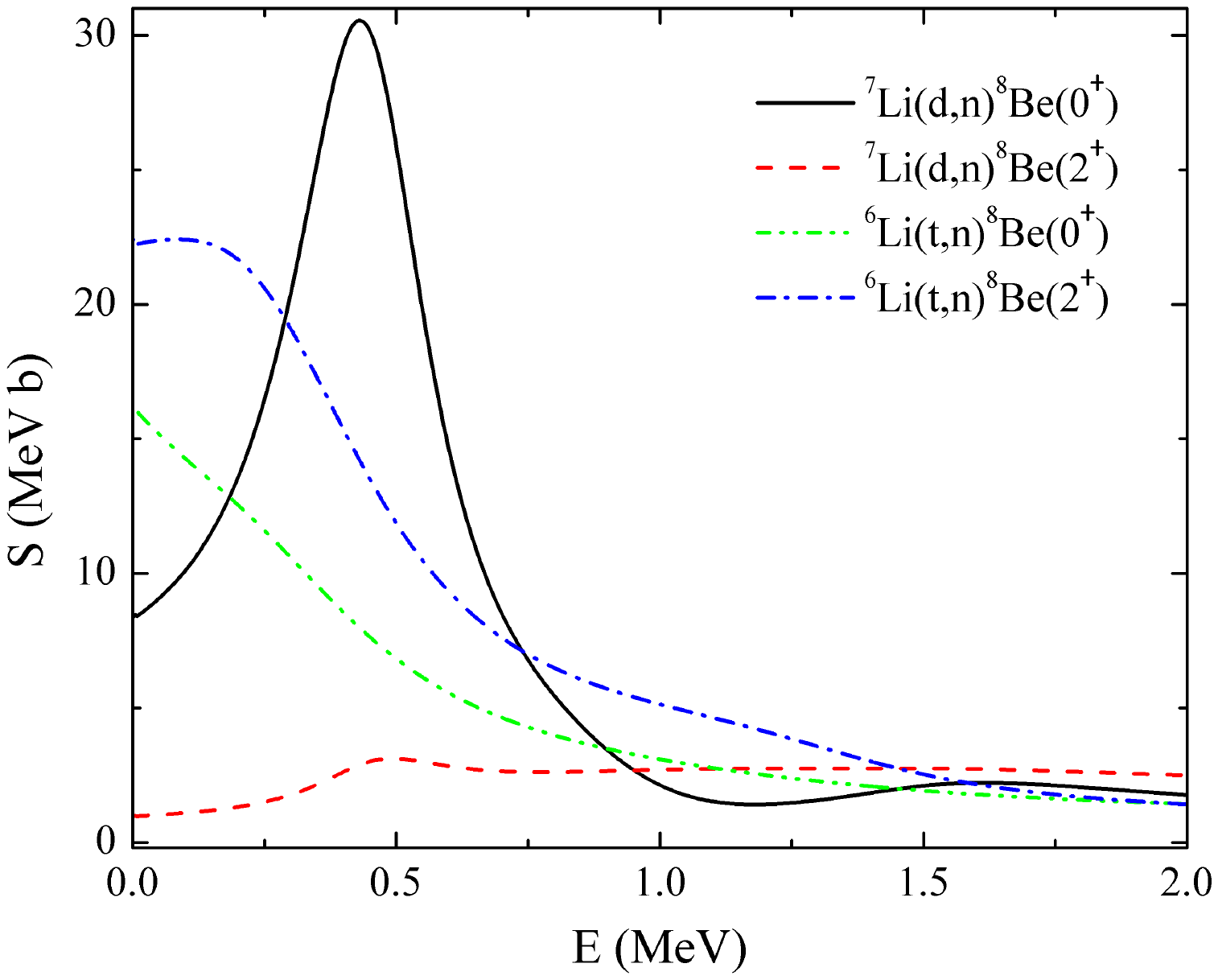}
[
\caption{Astrophysical S-factors of the reactions with the exit channel $^{8}$Be$+n$}
\label{Fig:Sfactors8Ben}%
\end{center}
\end{figure}

As demonstrated in Ref. \cite{PhysRevC.109.045803}, the astrophysical S-factors for the reactions 
$^{7}$Li$+d$ $\rightarrow$ $^{8}$Be($0^{+}$)$+n$ and $^{7}$Be$+d \rightarrow ^{8}$Be($0^{+}$)$+p$ are dominant when comparing reactions involving $^{7}$Li and deuterons, and $^{7}$Be with deuterons. It is insightful to compare these S-factors with those from reactions induced by $^{6}$Li+$^{3}$H and $^{6}$Li+$^{3}$He interactions, particularly those leading to the same exit channels. In Fig. \ref{Fig:Sfactors8Ben}, such a comparison is presented for the reactions leading to $^8$Be and a neutron in the exit channel. The S-factor of the reaction $^{6}$Li($^{3}$H,$n$)$^{8}$Be($0^{+}$) is higher than that of $^{7}$Li($d$,$n$)$^{8}$Be($0^{+}$) in the energy range 0$\leq E<$0.25 MeV. Conversely, in the energy range 0.25$<E<$0.9 MeV, the S-factor for the reaction $^{7}$Li($d$,n)$^{8}$Be($0^{+}$) is significantly dominant. Additionally, the S-factor for $^{6}$Li($^{3}$H,$n$)$^{8}$Be($2^{+}$) is the highest among all considered reactions at lower energies, 0$\leq E <$0.3 MeV. Summarizing, this indicates that $^{6}$Li burns more intensely than $^{7}$Li at energies 0$\leq E<$0.3 MeV, while $^{7}$Li consumption is more intense at energies 0.3$<E<$0.75 MeV.

\subsubsection{S-factors within Gamow windows for reactions  $^{6}$Li+$^{3}$H and $^{6}$Li+$^{3}$He}

We analyze the Gamow windows for the entrance channels $^{6}$Li+$^{3}$H and $^{6}$Li+$^{3}$He of reactions at a temperature $T_{9} = 0.8$ GK, as previously explored in Refs. \cite{2005ApJ...630L.105A, 2019PhRvL.122r2701R} and discussed in our earlier work \cite{PhysRevC.109.045803}. This temperature precisely defines the center $\left(E_{0}\right)$ and width $\left(\Delta E_{0}\right)$ of the Gamow peak energy, according to Ref. \cite{1999NuPhA.656....3A}. For the $^{6}$Li+$^{3}$H interaction, the Gamow window parameters are: $E_{0} = 276$ keV and $\Delta E_{0} = 318$ keV. Similarly, for the $^{6}$Li+$^{3}$He entrance channel, the parameters are: $E_{0} = 437$ keV and $\Delta E_{0} = 401$ keV.

\begin{figure}[hptb]
\begin{center}
\includegraphics[width=\textwidth]{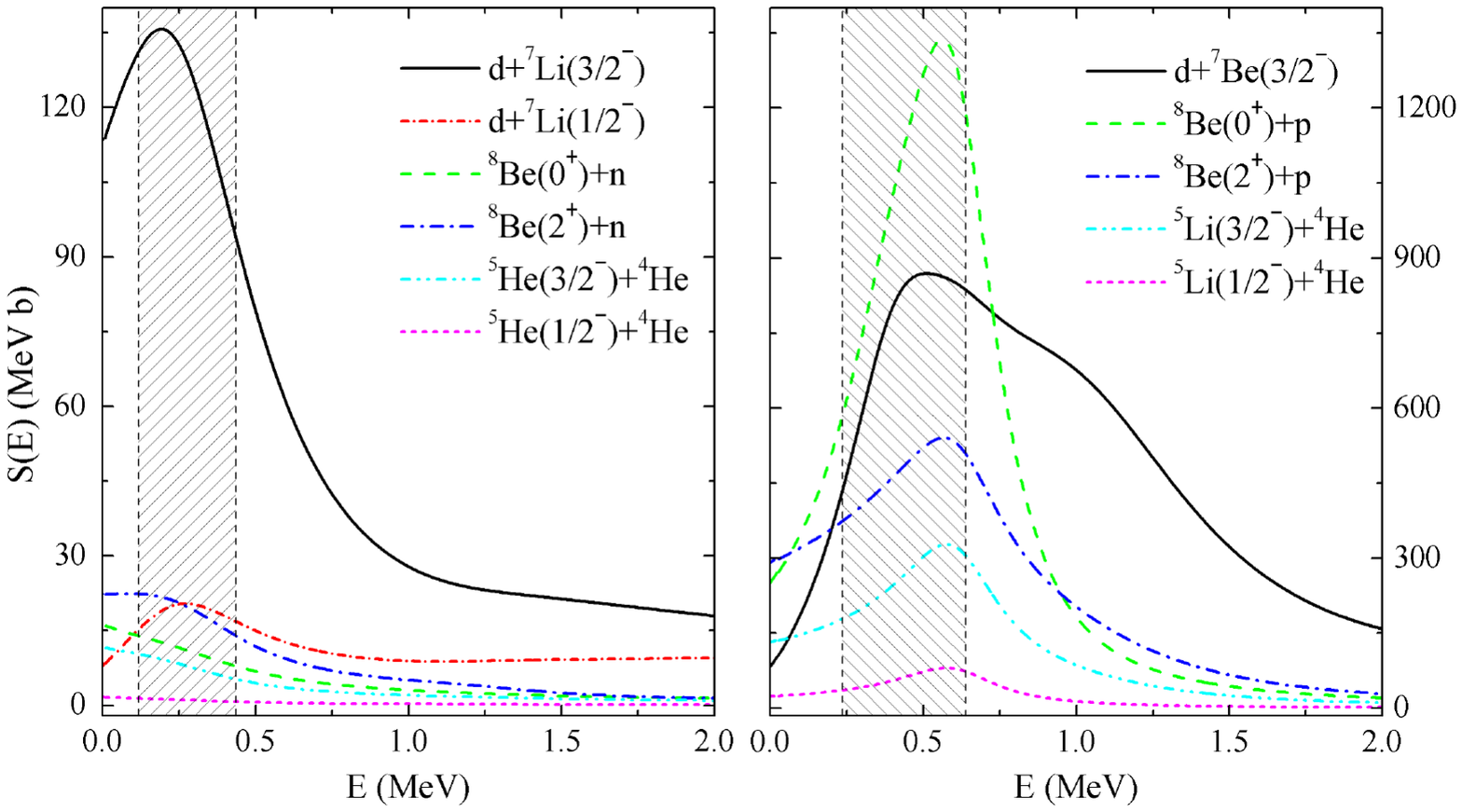}
\caption{Total astrophysical S-factors for reactions initiated by $^{6}$Li+$^{3}$H collisions (left panel) and $^{6}$Li+$^{3}$He collisions (right panel). The dashed areas represent the respective Gamow windows for each reaction.}
\label{Fig:Sfactors9Be9BGW}
\end{center}
\end{figure}

In Fig. \ref{Fig:Sfactors9Be9BGW}, we display the total astrophysical S-factors alongside the Gamow windows for reactions initiated by $^{6}$Li+$^{3}$H and $^{6}$Li+$^{3}$He collisions. This comparison highlights which reaction mechanisms are most influential within their respective Gamow windows for $^{9}$Be and $^{9}$B. Specifically, the S-factor for the reaction $^{6}$Li$(^{3}$He$,d)^{7}$Be(3/2$^{-}$) is found to be approximately seven times greater than that for $^{6}$Li$(^{3}$H$,d)^{7}$Li(3/2$^{-}$). Furthermore, the S-factor for the reaction $^{6}$Li+$^{3}$H $\rightarrow$ $^8$Be$(0^+)$ + $p$ surpasses that of its mirror reaction by an order of magnitude. In $^{9}$Be, the reaction $^{6}$Li$(^{3}$H$,d)^{7}$Li(3/2$^{-}$) is clearly the most significant within the Gamow window. Conversely, in $^{9}$B, the reactions $^{6}$Li$(^{3}$He$,d)^{7}$Be(3/2$^{-}$) and $^{6}$Li$(^{3}$He$,p)^{8}$Be(0$^{+}$) exhibit significant S-factors, indicating a complex interplay of reaction channels.

\begin{figure}[hptb]
\begin{center}
\includegraphics[width=\textwidth]{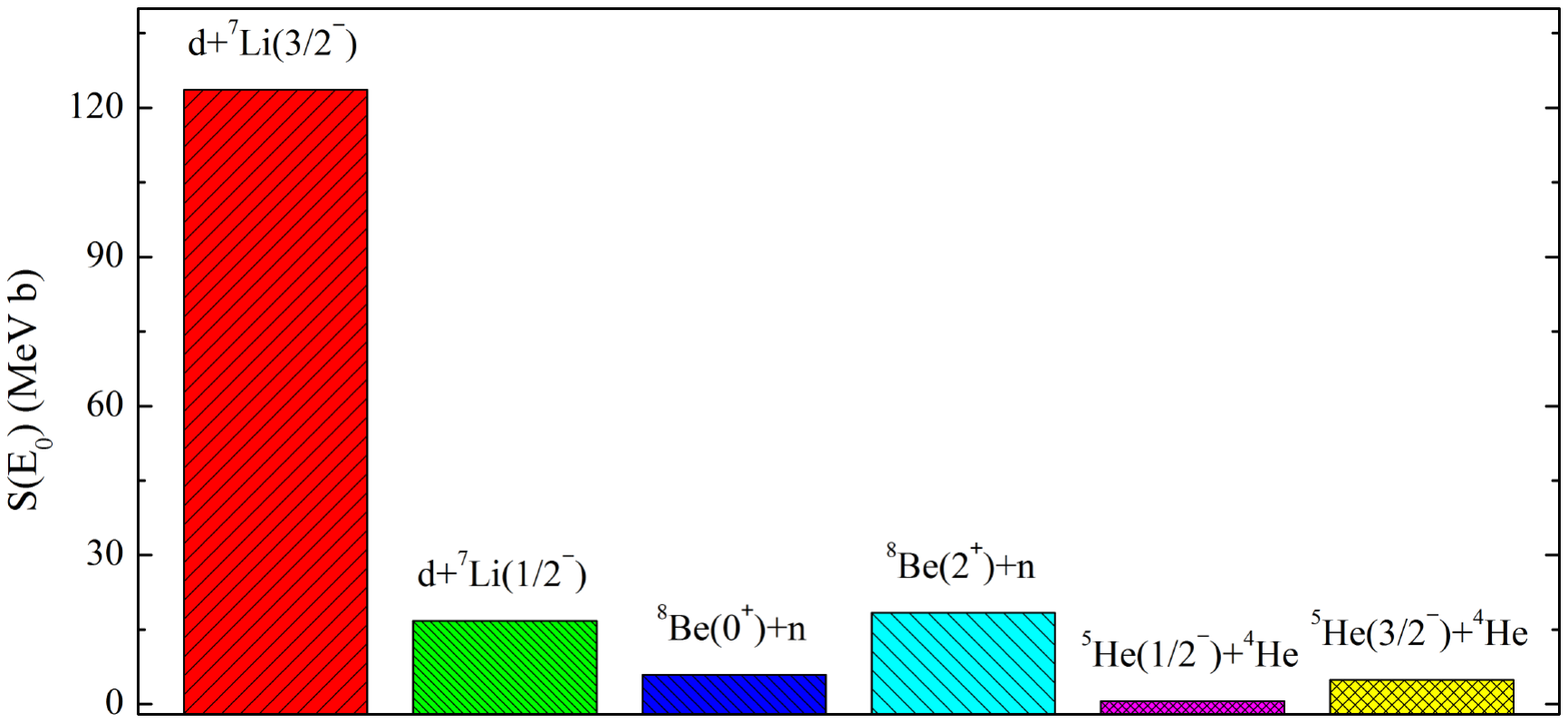}
\caption{Astrophysical S-factors of the reactions generated by $^{3}$H+$^{6}%
$Li collision at the Gamow peak energy $E_{0}$=276 keV}%
\label{Fig:Gamow9Be}%
\end{center}
\end{figure}

In Fig. \ref{Fig:Gamow9Be} we show the astrophysical S-factors of the
reactions initiated by the collision of $^{3}$H with $^{6}$Li. They are determined
at the Gamow energy $E_{0}$= 276 keV. The astrophysical S-factors of the
reaction $^{6}$Li$(^{3}$H$,d)^{7}$Li($3/2^{-}$), where the nucleus $^{7}$Li
 in the ground state is synthesized, overwhelmingly dominates over other reactions.

\begin{figure}[hptb]
\begin{center}
\includegraphics[width=\textwidth]{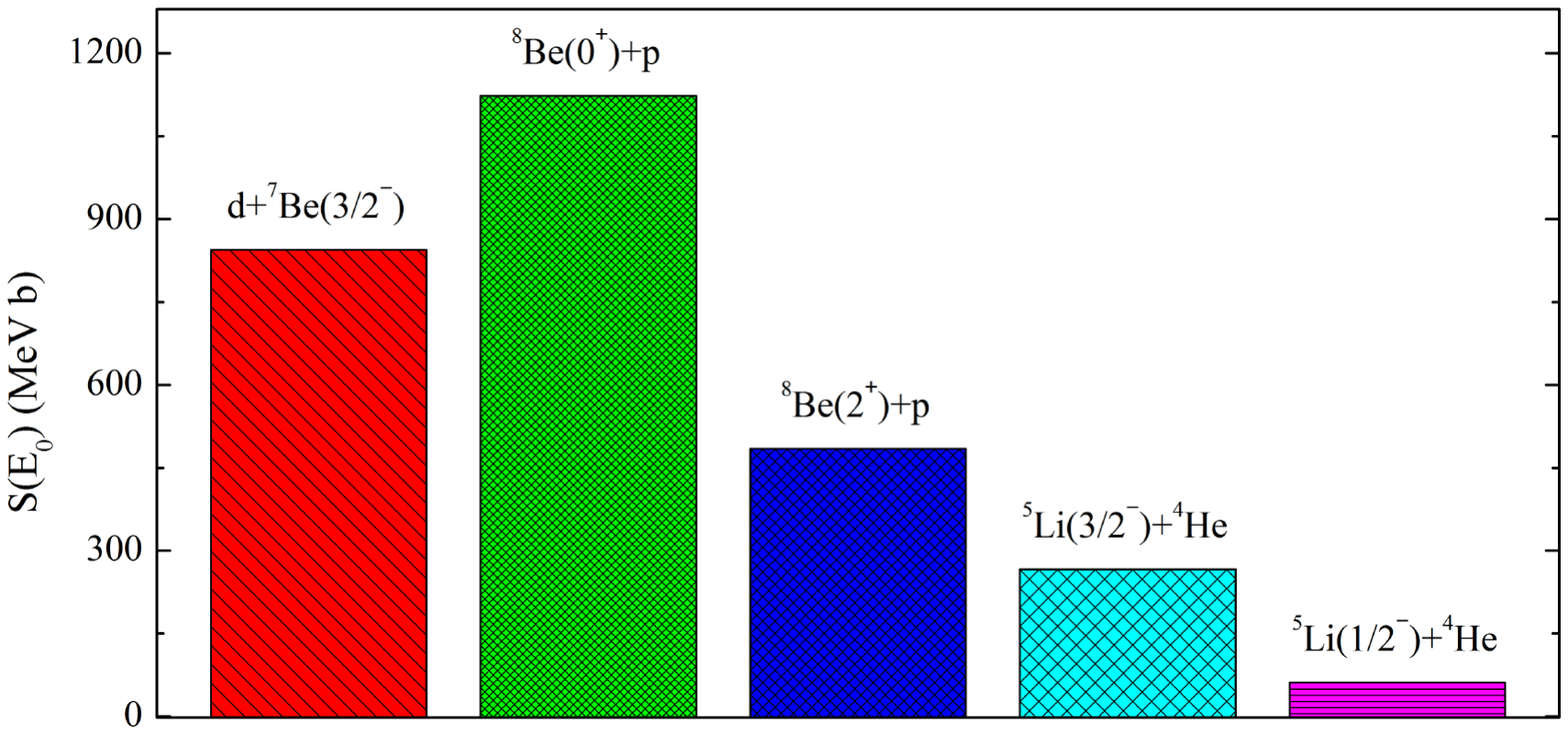}
\caption{The astrophysical S-factors of the reactions induced by $^{3}$He+$^{6}$Li collision at the Gamow energy $E_{0}$=437 keV.}
\label{Fig:Gamow9B}
\end{center}
\end{figure}

The astrophysical S-factors of the reactions initiated by the collision of $^{3}$He with $^{6}$Li at the Gamow energy $E_{0} = 437$ keV are displayed in Fig. \ref{Fig:Gamow9B}. At this energy, the reactions $^{6}$Li$(^{3}$He$, p)^{8}$Be($0^{+}$) and $^{6}$Li$(^{3}$He$, d)^{7}$Be($3/2^{-}$) are competitive (1123 and 884 MeV·b, respectively) and both dominate over other reactions.

\subsubsection{Theory versus Experiment}

As pointed out in the Introduction, there are no experimental evaluations of the astrophysical S-factors for the reactions considered in this paper. However, some measurements of total and differential cross sections at low energies are available. Eager to compare our results with all available experimental data, we decided  to estimate also the S-factors using  experimental data on differential cross sections. Assuming that all processes primarily proceed in s-wave at this energy range, the total cross section can be determined by multiplying the differential cross section by a factor of $4\pi$. From this, we straightforwardly obtained the desired values of the experimental S-factors.

\begin{figure}[ptb]
\begin{center}
\includegraphics[width=\textwidth]{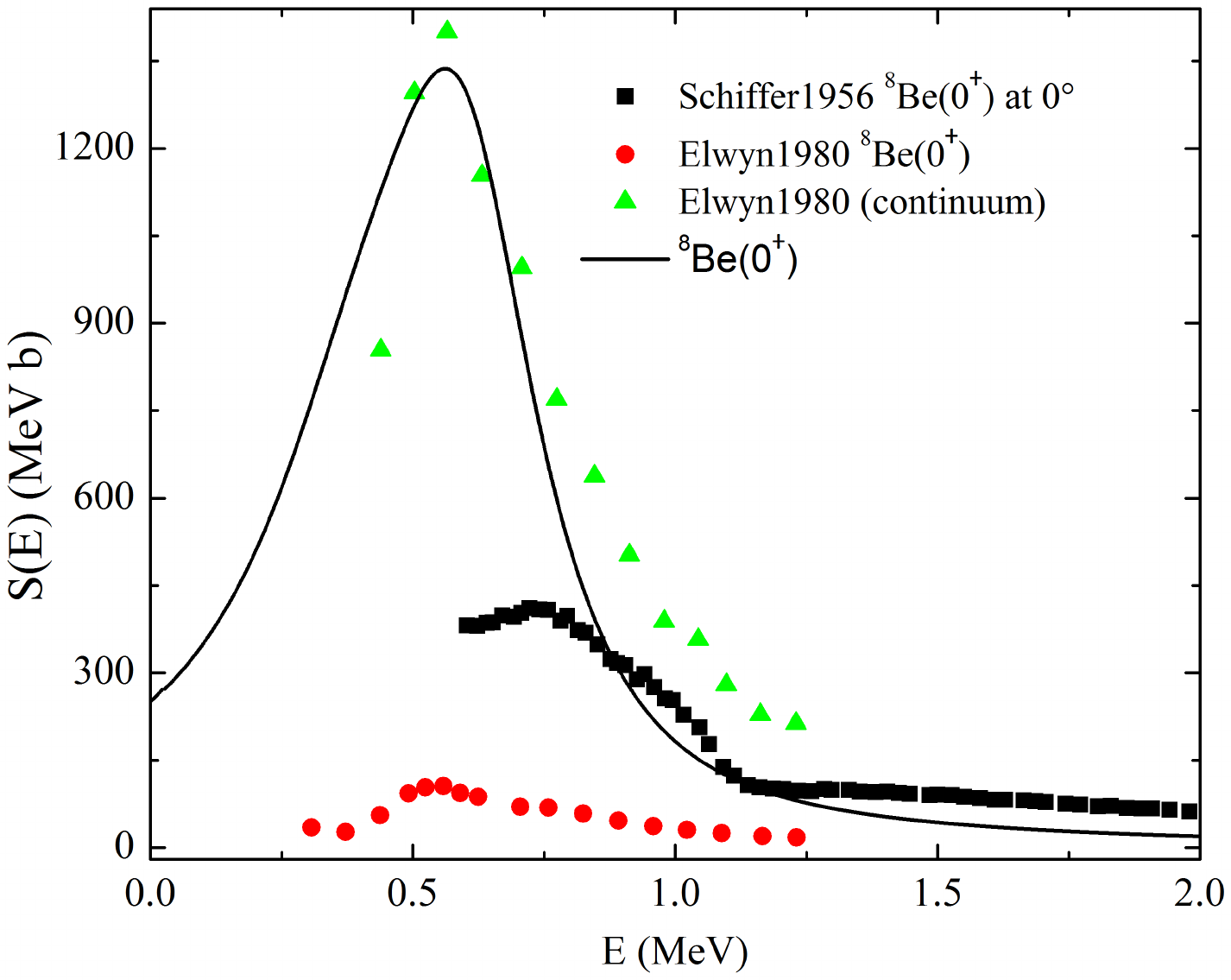}
\caption{Calculated and extracted experimental astrophysical S-factors of the reaction
$^{6}$Li+$^{3}$He=$^{8}$Be+$p$. Exit channels are shown near the curves.}%
\label{Fig:Sfactors9B6381}%
\end{center}
\end{figure}

Results of such evaluations are shown in Fig. \ref{Fig:Sfactors9B6381}, where experimental data are drawn from Ref. \cite{1956PhRv..104.1064S} (Schiffer1956), Ref. \cite{1980PhRvC..22.1406E} (Elwyn1980), and Ref. \cite{2021NIMPB.494...23Z} (Zhu2021). In Ref. \cite{1980PhRvC..22.1406E}, differential cross sections were determined for the exit channels with $^{8}$Be in the ground state, the first excited 2$^{+}$ state, and two high-energy resonance states at energies of 16.63 and 16.92 MeV. Additionally, a "continuum" cross section was obtained, which the authors defined to include protons from a broad 4$^{+}$ resonance state in $^{8}$Be, the sequential breakup of $^5$Li formed in the $^6$Li($^3$He, $\alpha$) reaction, and any direct three-body breakup contributions.
Interestingly, this S-factor is very close to our results for the exit channel $^{8}$Be($0^{+}$)+$p$. Above the energy range $E>$0.75 MeV, our results align closely with those obtained by Schiffer et al. \cite{1956PhRv..104.1064S}.

\begin{figure}[ptb]
\begin{center}
\includegraphics[width=\textwidth]{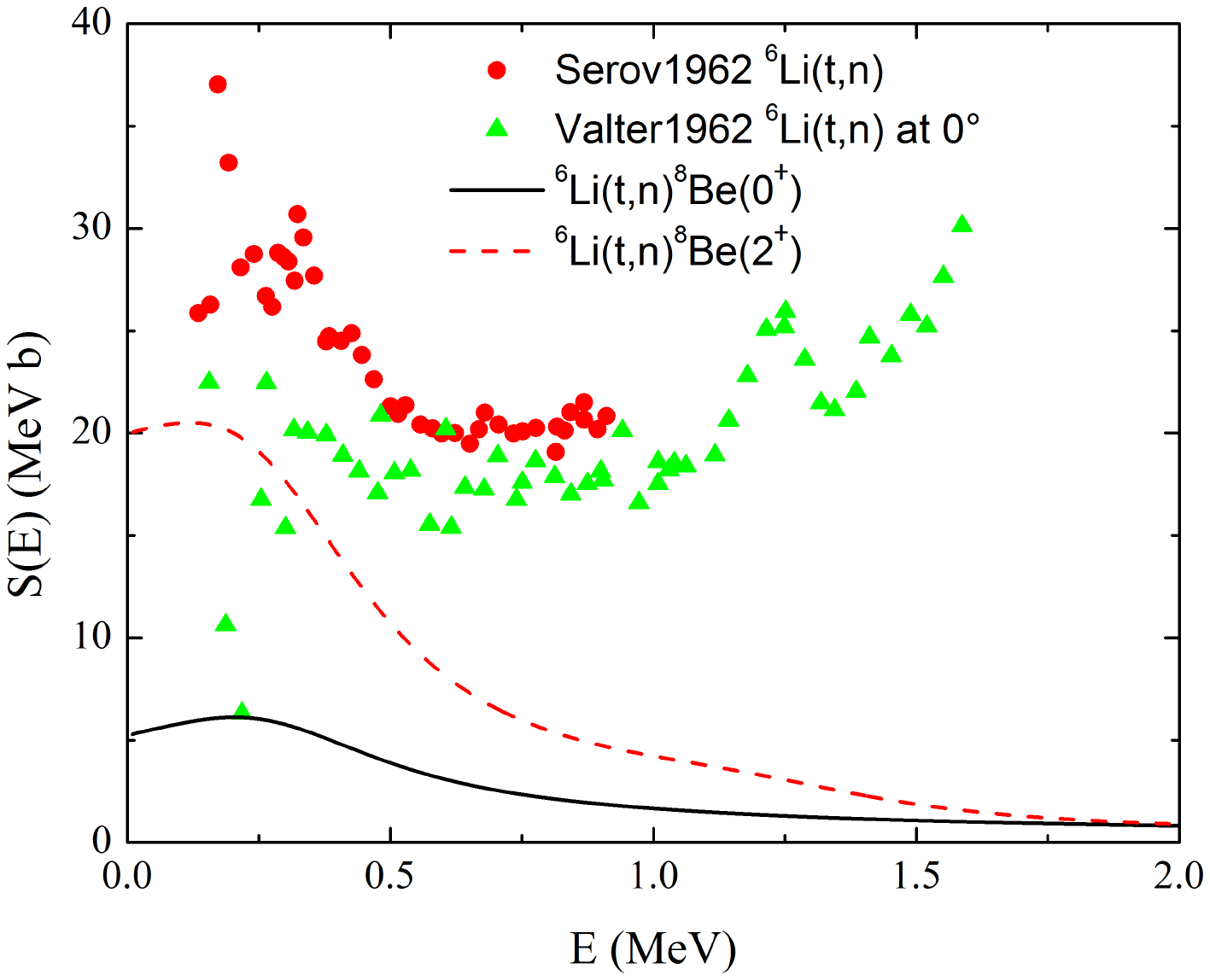}
\caption{Astrophysical S-factors of the reaction $^{6}$Li($^{3}$H,$n$). Exit channels are shown near the curves.}
\label{Fig:Sfactors9Be6381}%
\end{center}
\end{figure}

In Fig. \ref{Fig:Sfactors9Be6381}, we display astrophysical S-factors for the mirror reactions  $^{6}$Li+$^{3}$H=$^{8}$Be+$n$. Experimental data are extracted from Ref. \cite{Serov1962} (Serov) and Ref. \cite{Valter1962} (Valter). Our S-factor for the reaction $^{6}$Li+$^{3}$H=$n+^{8}$Be(2$^{+}$) is close to the S-factors extracted from the paper by Valter et al., as cited in Ref. \cite{Valter1962}.

\section{Conclusion}

We employed a multi-channel microscopic model to analyze the astrophysical S-factors for reactions initiated by $^6$Li when interacting with tritons:
\begin{equation}
^{6}\text{Li}(^{3}\text{H},d)^{7}\text{Li},\quad ^{6}\text{Li}(^{3}\text{H},n)^{8}\text{Be},\quad ^{6}\text{Li}(^{3}\text{H},\alpha)^{5}\text{He}
\label{eq:801}%
\end{equation}
and with $^3$He:
\begin{equation}
^{6}\text{Li}(^{3}\text{He},d)^{7}\text{Be},\quad ^{6}\text{Li}(^{3}\text{He},p)^{8}\text{Be},\quad ^{6}\text{Li}(^{3}\text{He},\alpha)^{5}\text{Li}.
\label{eq:802}%
\end{equation}
The analysis particularly focused on the formation of $^7$Li in the reaction $^6$Li($^3$H,d)$^7$Li, highlighting the nucleosynthesis of $^7$Li, while exploring how $^6$Li burns in the other listed reactions.

To align with their experimentally observed cluster structures, we treat the weakly bound nuclei $^{6}$Li, $^{7}$Li, and $^{7}$Be as two-cluster systems composed of $^{4}$He paired with $d$, $^{3}$H, and $^{3}$He, respectively. Additionally, the nuclei $^{5}$He, $^{5}$Li, and $^{8}$Be, which do not exhibit bound states and appear as resonance states, are also modeled as two-cluster systems consisting of $^{4}$He with $n$, $p$, and another $^{4}$He. These resonance states are approximated within our model as pseudo-bound states, facilitating a more manageable analysis of their behavior and interactions.

We have observed a significant asymmetry between the reactions initiated by $^{6}$Li collisions with $^{3}$H and those initiated by collisions with $^{3}$He. This asymmetry arises primarily from the Coulomb interaction and the differing relative positions of open and closed channels in $^{9}$Be and $^{9}$B. While the relative channel positions are influenced by the Coulomb force, they are predominantly determined by the central and spin-orbit components of the nucleon-nucleon potential. Consequently, a varied number of exit channels are active in the reactions occurring in $^{9}$Be compared to those in $^{9}$B.

\begin{acknowledgments}
This work received partial support from the Program of Fundamental Research of the Physics and Astronomy Department of the National Academy of Sciences of Ukraine (Project No. 0122U000889). We extend our gratitude to the Simons Foundation for their financial support (Award ID: 1290598). Additionally, Y.L. acknowledges the National Institute for Nuclear Physics, Italy, for providing a research grant to support Ukrainian scientists.  
\end{acknowledgments}

\bibliography{6Li3H}

\end{document}